\documentclass[twocolumn,showpacs,showkeys,preprintnumbers,amsmath,amssymb]{revtex4}

\usepackage{graphicx}
\usepackage{dcolumn}
\usepackage{bm}
\usepackage{wasysym}
\usepackage[dvips]{color}

\input{epsf}

\begin{document}

\preprint{Phys.Rev.B}

\title{The effect of electron--hole scattering on the transport properties of a 2D semimetal in a HgTe quantum well}

\author{ M.V. Entin$^{1}$, L.I. Magarill$^{1,2}$, E.B. Olshanetsky$^{1}$, Z.D. Kvon$^{1,2}$, N.N. Mikhailov$^{1}$,
S.A. Dvoretsky$^{1}$}
\affiliation{$^{1}$Institute of Semiconductor Physics, Siberian
Branch of the Russian Academy of Sciences, Novosibirsk, 630090, Russia
\\ $^{2}$ Novosibirsk State University, Novosibirsk, 630090, Russia}

\begin{abstract}
The influence of e-h scattering on the conductivity and
magnetotransport of 2D semimetallic HgTe is studied both
theoretically and experimentally. The presence of e-h scattering
leads to the friction between electron and holes resulting in a large
temperature-dependent contribution to the transport coefficients.
The coefficient of friction between electrons and holes is
determined. The comparison of experimental data with the theory
shows that the interaction between electrons and holes based on
the long - range Coulomb potential strongly underestimates the e-h
friction. The experimental results are in agreement with the model
of strong short-range e-h interaction.
\end{abstract}
\maketitle

\subsection*{Introduction}

Recently a 2D semimetal has been shown to be present in undoped
18-21 nm HgTe quantum wells with an inverted energy spectrum and
various surface orientations (013), (112) and (100)
\cite{hgte013,hgte112,hgte100}. It has been shown that this
semimetallic state is due to the overlap by an order of several
meVs of the conduction band minimum in the center of the Brillouin
zone and the valence band several maxima (the exact number and
configuration depending on the well surface orientation) situated
at some distance away from the Brillouin zone center. The Fermi
energy residing inside the energy interval corresponding to this
overlap results in a simultaneous existence of 2D electrons and
holes in the QW. The technology of low-temperature growth of a
composite ($SiO_2/Si_3N_4$) dielectric layer on top of the QWs has
allowed the fabrication of electrostatic top gate. Using this gate
makes it possible to obtain and study 2D semimetal states with any
desired ratio of electron and hole densities. The study conducted
in (013)-oriented HgTe wells has revealed certain features which
are peculiar to the transport in a 2D semimetal and may be
attributed to the electron--hole scattering inside the QW
\cite{e-h_scatt}. The present work presents a detailed theoretical
and experimental study of electron--hole scattering in a 2D
semimetal.

It is well known that in monopolar systems the interelectron
scattering does not affect the low-field conductivity. The
scattering between particles of the same kind conserves the total
momentum of the system. The momentum generated by an external
electric field does not dissipate, unless the impurities or
phonons are involved. As a result, in a system with a
simple electronic spectrum the conductivity does not depend on the
electron--electron scattering. This is not the case in a
multi-component system \cite{levin}, \cite{levin1}. In the absence of
mutual collisions, the components drift in an external electric field
with different velocities. The scattering between particles of
different sorts leads to the additional friction in the whole
system. In semimetal,  electrons and holes are accelerated by the
electric field in the opposite directions and the collision
between particles slowers the motion of both electrons and holes.
At low temperature the e-h scattering is limited (both for
electrons and holes) to the $kT$ interval near the Fermi surface
resulting in the temperature dependence of the probability of e-h
scattering  and of the corresponding corrections to
conductivity $\propto T^2$ \cite{e-h_scatt}. (However, in systems
with degenerate spectrum, e.g., in a 2D system in quantizing
magnetic fields the e-h scattering is not frozen out down to zero
temperature, see \cite{ent}). The paper is organized as follows.
Section I contains the theory of electron--hole scattering in a 2D
semimetal. Section II deals with the experimental details. In
Section III the comparison between the theory and experiment is
discussed.

\section{The theory of electron--hole scattering}
\subsection*{Kinetic equation solution}

\begin{figure}\centerline{\epsfxsize=10cm \epsfbox{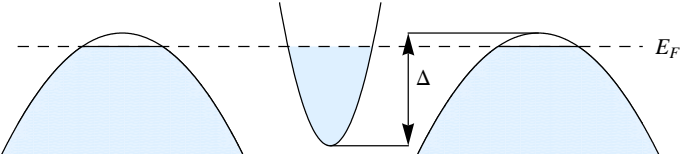} }
\caption{The energy band structure in a 20 nm (013) HgTe quantum
well.\label{fig11}}
\end{figure}

We consider a 2D semimetal with the $g_e$ equivalent electron valleys
and $g_h$ equivalent hole valleys centered in points ${\bf
p}_{e,i}$ and  ${\bf p}_{h,i}$, correspondingly. In particular, as
will be discussed in the next section, for the (013) HgTe QW
studied in the experiment we have a single conductance band valley
in the center of the Brillouin zone and two valence band valleys
situated along the [0$\bar{3}$1] direction, Fig.\ref{fig11}. The
conduction bands with energy spectra $\varepsilon_{\bf p}^e=({\bf
p}-{\bf p}_{e,i})^2/2m_e$ overlaps with  valence bands
$E_g-\varepsilon_{{\bf p}-{\bf p}_{h,i}}$, $\varepsilon_{\bf
p}^h=p^2/2m_h$ ($E_g>0$). Hole mass $m_h$ is assumed  to be
much larger than the electron mass $m_e$. The distances between
electron and hole extrema $|{\bf p}_{h,i}-{\bf p}_{e,j}|$ are
supposed to be large to suppress the electron--hole recombination.
At the same time, the scattering between electrons and holes
changing momenta near the extrema is permitted. Without the loss
of generality, further we will count the momenta from the band
extrema and replace ${\bf p}-{\bf p}_{h,j}\to {\bf p}$, ${\bf
p}-{\bf p}_{e,i}\to {\bf p}$.

The system of kinetic equations for the electron and hole
distribution functions $ f_{\bf p}^{e,h}$ reads

\begin{eqnarray}\label{kineq}
e_\nu{\bf E}\nabla_{\bf p} f_{\bf p}^{\nu}+[{\bf
p},\bm{\omega}_\nu]\nabla_{\bf p} f_{\bf
p}^\nu=\sum_{\nu'}I_{\nu,\nu'}+J_{\nu},
\end{eqnarray}
where  index $\nu=(+,-)$ numerates holes (h) and electrons (e),
respectively, $e_\pm=\pm e$, $-e$ is the electron charge, $
J_{\nu}$ is the collision integral of holes (electrons ) with
impurities, $I_{\nu,\nu'}$ are the inter-particle collision
integrals, $\bm{\omega}_{\nu}=e_\nu{\bf H}/m_{\nu}c$,
$\omega_{\nu}$ are cyclotron frequencies. In the linear
conductivity problem the collision integrals for particles of the
same sort ($I_{\nu,\nu}$) give no contribution to the
conductivity. Hence, the summation over $\nu'$ can be omitted
replacing $\nu'$ by $\bar{\nu}=-\nu$.

The hole-electron collision integral has the form
\begin{widetext}
\begin{eqnarray}\label{Ihe}\nonumber
    I_{he}=\frac{2\pi}{S^2}2g_e \sum_{{\bf p}',{\bf
    q},{\bf k},{\bf }'}|u_{\bf q}|^2\delta_{{\bf p}',{\bf p}+{\bf
q}}\delta_{{\bf k}',{\bf k}+{\bf q}}\delta(\varepsilon_{\bf
p}^h-\varepsilon_{{\bf
    p}'}^h+\varepsilon_{{\bf k}'}^e-\varepsilon_{\bf k}^e)\times\\
    \left[f_{\bf p}^h(1-f_{\bf p'}^h)f_{{\bf k}'}^e (1-f_{\bf k}^e)- f_{\bf p'}^h(1-f_{\bf
p}^h)f_{\bf k}^e(1-f_{{\bf k}'}^e)\right].
\end{eqnarray}
\end{widetext}
Here $u_{\bf q}$ is the Fourier transform of the electron--hole
interaction potential $u({\bf r})$, $S$ is the system area. Value $I_{eh}$ can be obtained  from Eq.(\ref{Ihe}) by exchange
$e\leftrightarrows h.$

We shall study linear (in ${\bf E}$) transport. Introducing linear
corrections $\phi_{\bf p}^\nu$ to equilibrium distribution
functions $f_{\bf p}^{0,\nu}$ and linearizing the kinetic
equations, we obtain
\begin{equation}\label{kineq1}
e_\nu({\bf E}\nabla_{\bf p}) f_{\bf p}^{0\nu}+([{\bf p},{\bm
\omega}_\nu]\nabla_{\bf p}) \phi_{\bf p}^\nu=\delta
I_{\nu,\bar{\nu}}+\delta J_{\nu},
\end{equation}
where $\delta I_{\nu,\bar{\nu}}$ and $\delta J_{\nu}$ are
linearized collision integrals,
\begin{widetext}
\begin{eqnarray}\label{Ihe}\nonumber
    \delta I_{\nu,\bar{\nu}}=\frac{2\pi}{S^2}2g_{\bar{\nu}} \sum_{{\bf p}',{\bf
    q},{\bf k},{\bf }'}|u_{\bf q}|^2\delta_{{\bf p}',{\bf p}+{\bf
q}}\delta_{{\bf k}',{\bf k}+{\bf q}}\delta(\varepsilon_{\bf
p}^\nu-\varepsilon_{{\bf
    p}'}^\nu+\varepsilon_{{\bf k}'}^{\bar{\nu}}-\varepsilon_{\bf k}^{\bar{\nu}})\times \\
    \Bigl\{\phi_{\bf p}^{\nu}\bigl[(1-f_{\bf p'}^{0\nu})f_{{\bf k}'}^{0\bar{\nu}}(1-f_{\bf
    k}^{0\bar{\nu}}) +f_{\bf p'}^{0\nu}f_{\bf k}^{0\bar{\nu}}(1-f_{{\bf k}'}^{0\bar{\nu}})\bigr ]
    -\phi_{\bf p'}^{\nu}\bigl[f_{\bf p}^{0\nu}f_{{\bf k}'}^{0\bar{\nu}} (1-f_{\bf k}^{0\bar{\nu}})+
    (1-f_{\bf p}^{0\nu})f_{\bf k}^{\bar{\nu}0}(1-f_{{\bf k}'}^{\bar{\nu}0})\bigr]+\nonumber  \\
     \phi_{{\bf k}'}^{\bar{\nu}}\bigl[f_{\bf p}^{0\nu}(1-f_{\bf p'}^{0\nu}) (1-f_{\bf k}^{0\bar{\nu}})+f_{\bf p'}^{0\nu}(1-f_{\bf p}^{0\nu})f_{\bf k}^{0\bar{\nu}}\bigr]
 -\phi_{\bf k}^{\bar{\nu}}\bigl[(1-f_{\bf p'}^{0\nu})f_{{\bf k}'}^{0\bar{\nu}} f_{\bf p}^{0\nu}+f_{\bf p'}^{0\nu}(1-f_{\bf p}^{0\nu})(1-f_{{\bf k}'}^{0\bar{\nu}}) \bigr ]
        \Bigr\}.
\end{eqnarray}
\end{widetext}

The solution of the system of kinetic equations can be searched in
the form of $\phi_{\bf p}^\nu= {\bf A}^\nu(\varepsilon_{\bf p}) {\bf
p}$, ${\bf A}^\nu(\varepsilon_{\bf p})\propto {\bf E}$. This
substitution results in the system of integral equations for ${\bf
A}^\nu(\varepsilon_{\bf p})$. Instead of solving this system we
use approximation
\begin{equation}\label{approx}
   \phi_{\bf p}^\nu\approx -{\bf p}{\bf V}^\nu
\partial_{\varepsilon_{\bf p}^\nu}f^{(0\nu)}_{\bf p},
\end{equation}
 where ${\bf V}^\nu$ are  the average velocities of particles.
To find the values  ${\bf V}^\nu$ one should integrate the kinetic
equations with the momentum ${\bf p}$. The impurity collision term
gives the rate of irretrievable momentum loss. The h-e collision
term determines the rate of momentum transfer between holes and
electrons (the force between subsystems of holes and electrons)
\begin{equation}\label{force}
   {\bm f}=2g_h\sum_{\bf p}{\bf p}I_{he}.
\end{equation}
The considered procedure is equivalent to the algebraization of
the collision terms
$$\delta J_{\nu}=-\frac{\phi^\nu}{\tau_{\nu}},~~~\delta
I_{\nu,\bar{\nu}}=\frac{\phi^{\bar{\nu}}}{\tau_{\bar{\nu}\nu}}-\frac{\phi^\nu}{\tau_{\nu\bar{\nu}}},$$
where $\tau_{\nu}$ is the transport relaxation for elastic
scattering on impurities. Relaxation times $\tau_{he}$ and
$\tau_{eh}$ of interparticle scattering satisfy the relation
$$\frac{ m_h}{N_s\tau_{he}}=\frac{m_e}{P_s\tau_{eh}}=\eta.$$
Quantity $\eta$ can be considered as the coefficient of liquid
friction between the subsystems of holes and electrons: the force
between electrons and holes is $SN_sP_s\eta({\bf V}^e-{\bf V}^h)$.

As a result we come to the system of hydrodynamic equations
\cite{levin,levin1} for ${\bf V}^\nu$
\begin{equation}\label{Vnu}
\frac{e_\nu}{m_\nu}{\bf E}+[{\bf V}^\nu,{\bm\omega}_\nu ]
-\frac{{\bf V}^\nu}{\tau_\nu}-\eta\frac{n_{\bar{\nu}}}{m_\nu}({\bf
V}^\nu-{\bf V}^{\bar{\nu}})=0.
\end{equation}
The system (\ref{Vnu})  can be written in the matrix form as
\begin{eqnarray}\label{Vnumatr}
    \Omega\cdot V = {\cal E},
\end{eqnarray}
where
\begin{widetext}
\begin{eqnarray}\label{Vnumatr}
    V = (V_x^e, V_y^e, V_x^h, V_y^h), ~~ {\cal E} =e(E_x/m_e, E_y/m_e, -E_x/m_h,
    -E_y/m_h), \nonumber \\
\Omega=\left(\begin{array}{cccc}
  -(\frac{1}{\tau_{e}}+\frac{1}{\tau_{eh}}) & \omega_e & \frac{1}{\tau_{eh}} & 0 \\
  -\omega_e &-(\frac{1}{\tau_{e}} +\frac{1}{\tau_{eh}}) & 0 &\frac{1}{\tau_{eh}} \\
  \frac{1}{\tau_{he}} & 0 & -(\frac{1}{\tau_{h}}+\frac{1}{\tau_{he}}) & \omega_h \\
  0 &\frac{1}{\tau_{he}} & -\omega_h &-(\frac{1}{\tau_{h}} +\frac{1}{\tau_{he}}) \\
\end{array} \right)
\end{eqnarray}
\end{widetext}
With the use of the solution of Eq.(\ref{Vnumatr})  $ V =
\Omega^{-1}{\cal E}$ we obtain $j_x=e(V_x^hP_s-V_x^eN_s)$,
$j_y=e(V_y^hP_s-V_y^eN_s)$ and
\begin{widetext}
\begin{eqnarray}\label{Vnumatr1}
&&\sigma_{xx}=N_1/D,~~~~\sigma_{yx}=N_2/D,\\&&\nonumber
N_1=e^2\Big(m_e m_h \left(m_e P_s \tau _h \left(\tau _e^2
   \omega _e^2+1\right)+m_h N_s \tau _e
   \left(\tau _h^2 \omega _h^2+1\right)\right)+\eta  \Big(2 m_e m_h \tau _e \tau _h
   \big(\left(N_s-P_s\right){}^2+\\&&N_s P_s
   \left(\tau _e \tau _h \omega _e \omega
   _h+1\right)\big)+N_s P_s \left(m_h^2\tau
   _e^2
   \left(\tau _h^2 \omega _h^2+1\right) +m_e^2 \tau _h^2 \left(\tau _e^2 \omega
   _e^2+1\right)\right)\Big)+\nonumber\\\nonumber &&\eta ^2 (N_s-P_s)^2\tau _e \tau _h\left(m_h  P_s
   \tau _e+m_e N_s
    \tau _h\right)\Big),\\
  \nonumber &&N_2=-e^2 \Big(m_e m_h \left(m_h N_s \omega _e
   \left(\tau _h^2 \omega _h^2+1\right) \tau
   _e^2+m_e P_s \tau _h^2 \left(\tau _e^2
   \omega _e^2+1\right) \omega
   _h\right) +2 \eta  m_e m_h\tau _e\tau _h\times\\\nonumber&&
   \left(N_s-P_s\right)  \left(N_s \tau_e \omega _e-P_s \tau _h \omega _h\right)
   +\eta ^2\tau _e^2\tau _h^2
   \left(N_s-P_s\right){}^2
   \left(m_e N_s \omega _e+m_h P_s \omega
   _h\right) \Big),\\ \nonumber
&&D=m_e^2m_h^2(1+\omega_e^2\tau_e^2)(1+\omega_h^2\tau_h^2)+2\eta
m_em_h(m_eN_s\tau_h(1+\omega_e^2\tau_e^2)+m_hP_s\tau_e(1+\omega_h^2\tau_h^2))+
\nonumber \\\nonumber
&&\eta^2((m_hP_s\tau_e+m_eN_s\tau_h)^2+\tau_e^2\tau_h^2(m_eN_s\omega_e+m_hP_s\omega_e)^2).
\end{eqnarray}
\end{widetext}
For components of resistivity tensor one can write
\begin{equation}\label{rho}
   \rho_{xx}=\frac{N_1D}{N_1^2+N_2^2}, ~~~
   \rho_{xy}=\frac{N_2D}{N_1^2+N_2^2}.
\end{equation}

At zero magnetic field $\rho_{xy}=0$, and the
temperature-dependent correction to the resistivity are simplified
\begin{widetext}
\begin{equation}\label{deltarho}
   \delta\rho(T)/\rho(T=0)=\frac{m_h N_s \tau
   _e+m_e P_s \tau _h}{m_e m_h}\frac{
   m_e m_h+\eta
   \left(m_h P_s \tau _e+m_e
   N_s \tau
   _h\right)}{
   m_
   h N_s \tau _e+m_e P_s \tau
   _h+\eta  \tau _e \tau _h
   \left(P_s-N_s\right){}^2}-1
\end{equation}
\end{widetext}

\subsection*{Electron--hole relaxation time}
The mean force acting between electron and hole subsystems ${\bf
f}$, Eq.(\ref{force}), in the Born approximation is determined by
substitution of the distribution functions of Eq.(\ref{approx}) into
Eq.(\ref{Ihe}). We arrive at
\begin{widetext}
\begin{eqnarray}\label{force1}
  {\bf f}= \frac{2\pi g_h}{(4\pi^2)^4}\int d{\bf p}\int d{\bf p'}\int d{\bf k}\int d{\bf
  k'}|u_{{\bf p}-{\bf p'}}|^2\delta({\bf p'}-{\bf p}+{\bf k}-{\bf k'})\delta(\varepsilon_{{\bf p}}^h-
  \varepsilon_{{\bf p}'}^h+\varepsilon_{{\bf k}'}^e-\varepsilon_{{\bf
  k}}^e) \nonumber
  \\ \Bigl\{({\bf p}-{\bf p'},{\bf p}){\bf V}_h(-\partial_{\varepsilon_{\bf p}^h}f^{(0h)}_{\bf p})
  \Bigl[f^{(0e)}_{\bf k}(1-f^{(0e)}_{\bf k'})f^{(0h)}_{\bf p'}+f^{(0e)}_{\bf k'}(1-f^{(0e)}_{\bf k})(1-f^{(0h)}_{\bf
  p'})- \\
  ({\bf k}-{\bf k'},{\bf k}){\bf V}_e(-\partial_{\varepsilon_{\bf p}^e}f^{(0e)}_{\bf k})
  \Bigl[f^{(0h)}_{\bf p}(1-f^{(0h)}_{\bf p'})f^{(0e)}_{\bf k'}+f^{(0h)}_{\bf p'}(1-f^{(0h)}_{\bf p})(1-f^{(0e)}_{\bf k'}) \Bigr]  \Bigr\}
\end{eqnarray}
\end{widetext}
The integral over ${\bf p}$ can be presented as $\int d{\bf p}=m_h
\int d\varepsilon_{\bf p} \int d\varphi_{\bf p}$ and similarly for the integral over
other momenta. Calculating  integrals over energies in the low
temperature limit we obtain the following expression
 for  the mean free time between
collisions of holes with electrons
\begin{widetext}
\begin{eqnarray}\label{tauhe1}
\frac{1}{\tau_{he}}\equiv \frac{N_s \eta}{m_h}=\frac{T^2
m_e^2m_h}{(4\pi^2)^3}\frac{\zeta^2}{2P_s
\hbar^7}~\int_0^{2\pi}d\phi d\varphi d \varphi'(1-\cos{\phi})
\delta(\zeta(\cos{\phi}-1)+\cos{\varphi}-\cos{\varphi'})\times\nonumber
\\\delta(\zeta\sin{\phi}+\sin{\varphi}-\sin{\varphi'})
|u_{p_{Fh}(1-\cos{\phi})}|^2,~
\end{eqnarray}
\end{widetext}
where  $\zeta=p_{Fh}/p_{Fe}, ~ p_{Fh}, p_{Fe}$ are the Fermi
momenta of holes and electrons respectively.

Expression (\ref{tauhe1}) can be transformed to
\begin{eqnarray}\label{tauhe2}
\frac{1}{\tau_{he}}=\frac{m_e^2}{12\pi^3
g_h\hbar^5}\frac{T^2}{\epsilon_{Fh}}\zeta\int_0^{x_0}dx~~\frac{x
|u_{ 2p_{Fh}x}|^2}{\sqrt{1-x^2}\sqrt{1-\zeta^2x^2} }.
\end{eqnarray}
Here $x_0={\mbox{ min}(1,1/\zeta)}$.

The Fourier transform of the Coulomb e-h interaction $u_{\bf q}$
depends on the structure of the system and the screening. In the
simple 2D model of electron gas the Coulomb interaction with
linear screening reads as
\begin{equation}\label{Coulomb}
    u_q=\frac{2\pi e^2}{\chi}\frac{1}{q+\kappa},
\end{equation}
where the screening constant is collected from the individual
screening constants of electron and hole gases,
$\kappa=\kappa_e+\kappa_h=2(g_h/a_{B,h}+g_e/a_{B,e})$,
$a_{B,e}=\hbar^2\chi/m_he^2$ and $a_{B,h}=\hbar^2\chi/m_ee^2$ are
the Bohr radiuses of electrons and holes, respectively; $\chi$ is
the effective dielectric constant.

This expression is valid in the linear screening approximation
that needs smallness of $\kappa$ as compared to the transmitted
momentum min$(p_{F,e},p_{F,e})$. Besides, here we neglect the
width of the quantum well. As a result, the potential becomes
independent of the HgTe dielectric constant. This 2D consideration
loses applicability in the specific system under the consideration
where the quantum well width $d\gtrsim 1/\kappa$.  In fact, the
screening radius $1/\kappa$ should be limited from below by $d$.

Accounting for finite width of the quantum well leads to
replacement of the 2D potential by
\begin{equation}\label{Coulombcorrected}
    u_q=\frac{2\pi e^2}{\chi}\frac{F(qd)}{q+\kappa F(qd)},~~~~~q<2(p_{Fe},p_{Fh}).
\end{equation}
where the function $F(qd)$ follows from the solution of
electrostatic interaction problem   of two  singly charged
particles placed inside a layer of width $d$ between two
semi-infinite dielectrics. Using planar Fourier transform, we have
for the interaction of two  point charges located in the points
$z,z'$, $d/2>z> z' >- d/2$:
\begin{widetext}
\begin{equation}\label{22}
   -\frac{e^{-q (z+\text{z'})} \left(e^{d q} (r+1)-e^{2
   q z} (r-1)\right) \left(-r+e^{q (d+2 \text{z'})}
   (r+1)+1\right)}{2\chi q \left(e^{2 d q}
   (r+1)^2-(r-1)^2\right)}.
\end{equation}
\end{widetext}

For $z<z'$ it is necessary to do replacements $z \leftrightarrow
z'$ in Eq.(\ref{22}). Here $r=\chi/\chi_{\tiny \mbox{HgTe}}$,
$\chi $ is the dielectric constant of external layers ($CdTe$). To
find the function $F(x),$ one should integrate the potential
(\ref{22}) with the squares of electron and hole transversal wave
functions $|\psi_{e}(z)|^2$ and $|\psi_{h}(z')|^2$. For a well
with hard walls $\psi_{e,h}(z)=\sqrt{2/d}\cos(\pi z/d)$. In this
case, we find
\begin{widetext}
\begin{eqnarray}\label{23}
F(x)=\frac{r}{\left(x^2 +
       4\pi^2 \right)^2}\left(x\left (3 x^2 +
      20\pi^2 \right) + 32\pi^4\frac {\left (-1 + e^x \right) x +
      r\left (e^x (x - 2) + x + 2 \right) } {\left (r +
       e^x (r + 1) - 1 \right)x^2}\right)
\end{eqnarray}
\end{widetext}

Assembling the previous expressions we get the $\tau_{he}$ for the
Coulomb scattering
\begin{widetext}
\begin{eqnarray}\label{tauhe}
\frac{1}{\tau_{he}}=\frac{m_ee^4}{6\pi\chi^2g_h\hbar^3}\frac{m_e}{m_h}\frac{T^2}{\epsilon_{Fh}^2}\zeta\int_0^{x_0}dx~~\frac{x
F^2(w x)}{\sqrt{1-x^2}\sqrt{1-\zeta^2x^2}(2x+\xi F(wx))^2 },
\end{eqnarray}
\end{widetext}
 where
$w=2 p_{Fh}d, ~ \xi=\kappa/p_{Fh}$. In the strict 2D case $w=0$,
the potential converts to Eq.(\ref{Coulomb}) and  $F(wx)$ should
be replaced by 1. It should be emphasized that in the real case
the parameter $\xi$ is large, so, the mean free time ceases to
depend on $F$. This conclusion is valid in the linear screening
theory. Careful examination shows the necessity of revision of this
approach. Quantity $1/\tau_{he}$ is proportional to $T^2$.
This results in a similar temperature dependence of the correction
to the residual resistivity at low temperature.

\subsection*{Short-range interaction}
Together with the long-range Coulomb part the interaction between
electrons and holes contains also
 the short-range kernel interaction. Large dielectric constant of HgTe and CdHgTe leads
 to the dielectric screening of the Coulomb contribution. In that case the on-site e-h interaction  can  prevail. To
estimate the kernel contribution one can replace $u_{ 2p_{Fh}x}$
by a constant:
\begin{equation}\label{Lambda}
   u_{
2p_{Fh}x}=\pi\hbar^2\frac{m_e+m_h}{m_em_h}\Lambda; ~~~ \Lambda=\frac{m_em_h}{m_e+m_h}\frac{1}{\hbar^2\pi}\int
u(r)d{\bf r}.\end{equation}
 Dimensionless quantity $\Lambda$
describes the strength of  the contact e-e-interaction. As a result we find for $\eta:$
\begin{eqnarray}\label{tauhe11}
\eta=\frac{(m_e+m_h)^2\Lambda^2T^2}{24\pi^2\hbar^3N_sP_s}\ln|\frac{1+\zeta}{1-\zeta}|
\end{eqnarray}
In accordance with (\ref{tauhe1}), the model of isotropic energy
spectrum leads to a logarithmic divergency of the temperature
corrections to the conductivity at equal Fermi momenta of
electrons and holes. The divergency originates from the
 probability of two Fermi particle
backscattering with conservation of their individual energies. For
the isotropic Fermi surfaces such processes occur for all
electrons on the Fermi surface.

\subsection*{Anisotropic spectrum}
 In fact,
the holes in the system under consideration have  anisotropy.
That evidently limits kinematically the possibility of
backscattering and the divergency. This makes it necessary to take
into account  the hole spectrum anisotropy neglected before. It
can be done in the relaxation time approximation for elliptic hole
spectrum for the case of zero magnetic field. The anisotropy of
the spectrum results in the   anisotropy of  temperature
corrections. In accordance with the experimental situation, we
shall consider the electric field applied along the symmetry axes,
say $i$. In this case Eq.(\ref{Vnu}) is modified as
\begin{equation}\label{Vnu2}
\frac{e}{m_i}{ E_i} -\frac{ V_i^h}{\tau_h}-\eta\frac{P_s}{m_i}(
V_i^h- V_i^e)=0.
\end{equation}
Here, subscript $i$ marks the specific direction of the field
and the same component of the hole mass.

Friction coefficient $\eta_i$ for the same direction of
electric field reads
\begin{equation}\label{etai}
\eta_i=\frac{m_h}{N_s\tau_{he,i}}=\frac{T^2(m_e+m_h)^2
\Lambda^2}{6\pi^3\hbar^3N_sP_s} f(\alpha_i,\zeta),
\end{equation}
where $m_h=\sqrt{m_1m_2}$ is the mass of the hole density of
states, $\alpha_i=\sqrt{m_i/m_h}$,

\begin{widetext}
\begin{eqnarray}\label{f}
f(\alpha,\zeta)=\frac{(\alpha\zeta)^2}{32}\int_0^{2\pi} d\varphi
d\varphi' d\phi d\phi' (\cos
\varphi-\cos\varphi')\delta(\alpha\zeta(\cos
\varphi'-\cos\varphi)+\cos\phi-\cos\phi')\nonumber \\
\times
\delta((\zeta/\alpha)(\sin\varphi'-\sin\varphi)+\sin\phi-\sin\phi').
\end{eqnarray}
\end{widetext}

\begin{figure}[h]
\centerline{\epsfxsize=7cm \epsfbox{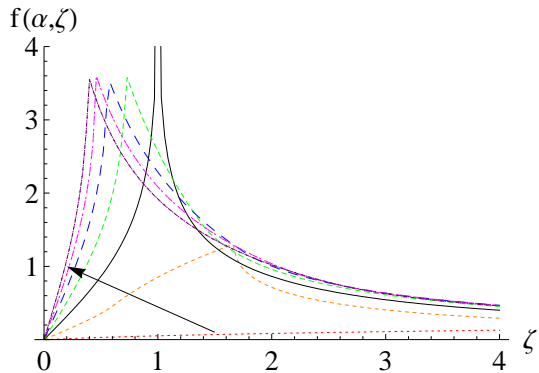} } \caption{Dependence
of temperature correction in the anisotropic case on the
electron--to--holes concentration ratio and the hole masses ratio
via parameters $\zeta$ and $\alpha$. Parameter $\alpha$
 runs values 0.2,0.6,1,1.4,1.8,2.2,2.6. The irection of $\alpha$ growth is shown by arrow. \label{figg2}}
\end{figure}

Integrals over three angles can be evaluated and we arrive at
\begin{widetext}
\begin{eqnarray}\label{f1}
f(\alpha,\zeta)= \alpha^4\int_0^{1}
\frac{x^2dx}{\sqrt{1-x^2}({1+x^2(\alpha^4-1)})} \ln
\Bigg|\frac{\alpha+\zeta\sqrt{1+x^2(\alpha^4-1)}}{\alpha-\zeta\sqrt{1+x^2(\alpha^4-1)}}\Bigg|.
\end{eqnarray}
\end{widetext}
Fig.\ref{figg2} shows the dependence of $f(\alpha,\zeta)$ on
$\zeta$ for different $\alpha$. All curves contain the limited
singularities corresponding to the equality of the hole Fermi
ellipses axes to the diameter of the electron Fermi circle. The
exception is the case of $\alpha=1$, when $f(1,\zeta)\propto
-\ln|\zeta-1|$ at $\zeta\to 1$. In this case, the divergency can be
limited by the finite temperature or collision widening.

\section{Experiment}

\subsection*{Samples}

\begin{figure}[h]
\begin{center}\leavevmode
\includegraphics[width=0.9\linewidth]{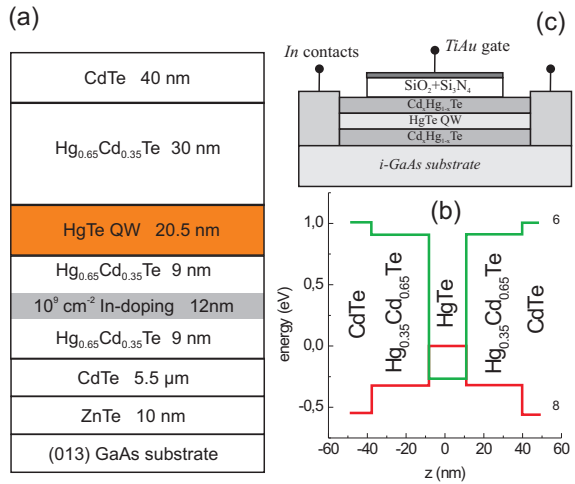}
\caption{The quantum well layer structure - (a), the quantum well
energy diagram -(b), and the cross section of the samples studied.
\label{Fig.1}}
\end{center}
\end{figure}

The $Hg_{0.3}Cd_{0.7}Te/HgTe/Hg_{0.3}Cd_{0.7}Te$ quantum wells
with the (013) surface and the  thickness of 20.5 nm were prepared by
molecular beam epitaxy. The details of the structure growth
process are described in \cite{technology,sassine}. The QW
cross-section and the energy diagram of the structures
investigated is shown in Fig.\ref{Fig.1}a and 1b, respectively. To
perform magnetotransport measurements, the samples based on these
quantum wells were prepared by standard photolithography in the
form of 50 $\mu$m wide Hall bars with the voltage probes spaced
100 $\mu$m apart. The ohmic contacts to the two-dimensional gas
were formed by the in-burning of indium. To change and control the
electron and hole densities in the QW, the electrostatic top gate
has been supplied. For this purpose, a dielectric layer containing
100 nm $SiO_2$ and 200 nm $Si_3N_4$ was first grown on the
structure using the plasma-chemical method. Then, the TiAu gate
was deposited. The schematic drawing of the devices prepared in
this way is shown in Fig.\ref{Fig.1}c. The magnetotransport
measurements in the described structures were performed in the
temperature range of 0.2--4.1 K in magnetic fields up to 5 T by the
standard four-point circuit at the 12--13 Hz ac signal with the
current of 1--10 nA through the sample, which is sufficiently low
to avoid the overheating effects.

\subsection*{Experimental results}

\begin{figure}[h]
\begin{center}\leavevmode
\includegraphics[width=0.9\linewidth]{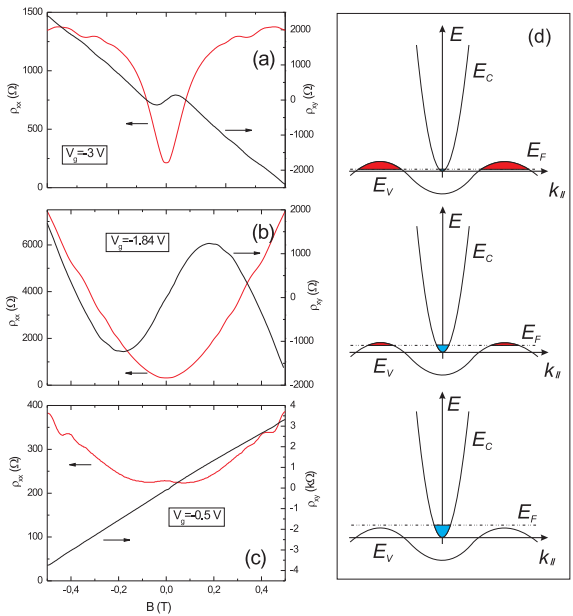}
\caption{Magnetic field dependences $\rho_{xx}(B)$ and
$\rho_{xy}(B)$ for the 2D electron-–hole system in the HgTe
quantum well at $T = 0.19$\,K for three gate voltages: (a)-
$V_{g}=-3$ V; (b)- $V_{g}=-1.84$ V and (c)-$V_{g}=-0.5$ V; (d)-
the energy band diagrams with approximate positions of the Fermi
energy corresponding to the curves on the left side.
\label{Fig.2}}
\end{center}
\end{figure}

To gather information about the structures properties and to
determine the main transport parameters of the system
corresponding to different gate voltages, the magnetic field
dependences of the diagonal $\rho_{xx}(B)$ and Hall
$\rho_{xy}(B)$ components of the resistance tensor were measured.
These functions show a strong dependence on the magnitude and sign
of the gate voltage applied to the sample. Fig.\ref{Fig.2}a,b,c
present the curves measured at gate voltages -3, -1.84 and
-0.5 V respectively. One can see that an alternating-sign Hall
effect and strong positive magnetoresistance are observed at $V_g$
= –3 and -1.84 V (see Fig.\ref{Fig.2}a,b). Meanwhile, at $V_g$ =
-0.5 V (see Fig.\ref{Fig.2}c), there is a weak negative
magnetoresistance at low fields and positive magnetoresistance at
higher fields, and the magnetic field dependence of the Hall
resistance is linear with its slope opposite to that of
$\rho_{xy}(B)$ at $V_g$ = –3 and -1.84 V and $\vert $B$\vert >$0.1
T and $\vert $B$\vert >$0.4 T, respectively.

\begin{figure}[h]
\begin{center}\leavevmode
\includegraphics[width=0.9\linewidth]{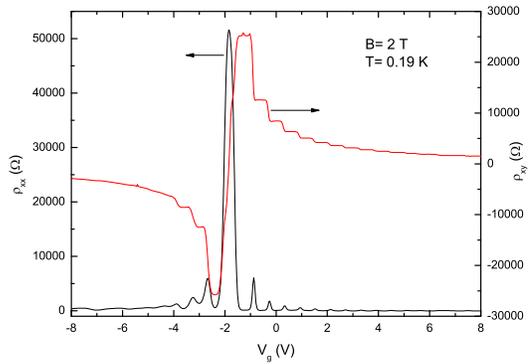}
\caption{Gate voltage dependences $\rho_{xx}(B)$ and
 $\rho_{xy}(B)$ for the 2D electron-–hole system in the HgTe quantum
 well at $T = 0.19$\,K and magnetic field $B=2$ T. \label{Fig.3}}
\end{center}
 \end{figure}

The described behavior suggests that, by varying the gate voltage
we change the carrier type content in the quantum well. This
conclusion is further supported by the $\rho_{xx}(V_g)$ and
$\rho_{xy}(V_g)$ traces measured at a constant magnetic field B=2
T corresponding to the quantum Hall effect regime,
Fig.\ref{Fig.3}. The quantum Hall plateaux in $\rho_{xy}(V_g)$ and
minima in $\rho_{xx}(V_g)$ are well developed for filling factors
$\nu=1-10$ on the electron side and for $\nu=1-4$ on the hole side
indicating a very high quality of the samples investigated. At
$V_g\approx-1.8$V a dramatic change of the sign of $\rho_{xy}$ takes
place signifying a change of the predominant carrier type in the
well. This transformation in $\rho_{xy}(V_g)$ is accompanied by a
sharp peak in $\rho_{xx}(V_g)$. The behavior of our system in the
quantum Hall effect regime has been studied earlier \cite{hall}.

\begin{figure}[h]
 \begin{center}\leavevmode
\includegraphics[width=0.9\linewidth]{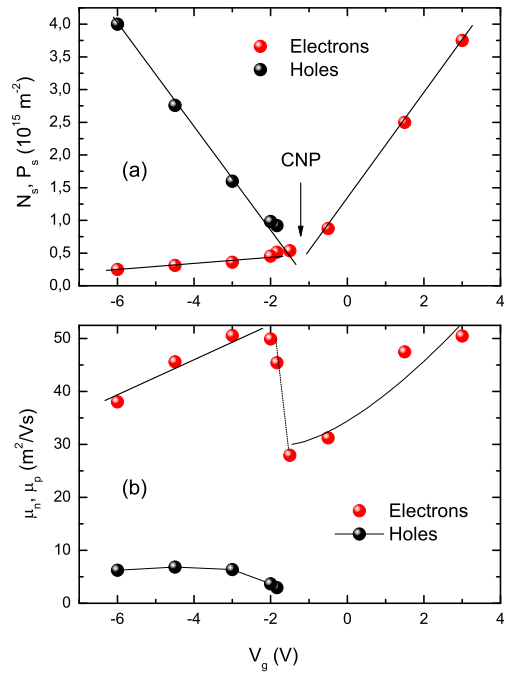}
\caption{(a)- the electron $N_s$ and hole $P_s$ densities versus
 gate voltage; (b)- the electron $\mu_n$ and hole $\mu_p$
 mobilities versus gate voltage; $T = 0.19$ K. \label{Fig.4}}
\end{center}
 \end{figure}

By fitting the dependences similar to those presented in
Fig.\ref{Fig.2}a,b,c using the formulas of the standard classical
transport model in the presence of two groups of carriers of
opposite signs \cite{Bonch}, we can determine the types of charge
carriers involved in the  transport, as well as their mobilities and
densities. Fig.\ref{Fig.4} presents these parameters as functions
of the gate voltage. We first consider the gate voltage
dependences of the electron and hole densities shown in
Fig.\ref{Fig.4}a. For $V_g\ge -1$ V, the experimental curves (see,
e.g., Fig.\ref{Fig.2}c) are adequately described by the transport
model involving only electrons as charge carriers. Although holes
can be present in this case with a density much lower than the
electron density, their contribution to the transport is immaterial
due to their lower mobility. As would be expected, the gate
voltage dependence of the electron density is linear with a slope
of $8.12\times 10^{14} m^{-2}V^{-1}$ corresponding to the
capacitance of the dielectric. An absolutely different pattern is
observed for $V_g \le -1.5$ V. To describe the dependences
similar to those presented in Fig.\ref{Fig.2}a,b two types of
carriers, electrons and holes, should be taken into account.
Fig.\ref{Fig.4}a shows the electron and hole densities as
functions of the gate voltage for $V_g \le -1.5$ V obtained from
the processing of the experimental data. Clearly, as the negative
gate bias increases, the hole density increases and the electron
density decreases linearly with slopes $7.9\times10^{14}$ and
$0.7\times10^{14}$ $m^{–2}$ $V^{–1}$, respectively. We note that
the sum of the magnitudes of these slopes is about the magnitude
of the slope of $N_s(V_g)$ for $V_g\ge -1$ V, as would be
expected, because electrons are the only observable type of
carriers for $V_g\ge -1$ V. Moreover, the slope ratio
$P_s(V_g)/N_s(V_g)\approx 11.3$ for $V_g \le -1.5$ V should
correspond to the ratio of the densities of states of holes and
electrons. Then, if holes fill two valleys (as expected for a
(013) 20 nm HgTe QW) and electrons fill only one valley, then the
hole mass is $m_h \approx 0.15 m_0$ if we take the electron mass
$m_e \approx 0.025 m_0$. These values are close to those
determined from the cyclotron resonance measurements
\cite{cyclotron}.

Processing diagonal $\rho_{xx}(B)$ and Hall $\rho_{xy}(B)$
dependences in the vicinity of the gate voltages where the
electron and hole densities are close is rather difficult.
However, extrapolating  linear dependences $N_s(V_g)$ and
$P_s(V_g)$, we find their crossing point, where the electron and
hole densities are equal, $V_g\approx -1.3$ V - the so called
charge neutrality point (CNP).

Using the electron and hole density versus gate voltage
dependences obtained above we can plot the opposite to
Fig.\ref{Fig.2}a,b,c qualitative energy band diagrams with the
corresponding Fermi level positions and the conduction and valence
band occupation, Fig.\ref{Fig.2}d. As mentioned above for the
(013) HgTe QWs we have a single conductance band minimum in the
center of the Brillouin zone and two valence band maxima situated
along the [0-13] direction. According to the CNP electron and hole
densities and their mass values determined above the conductance
and the valence bands overlap in our samples is about 10 meV.

Now, we consider the behavior of the electron and hole mobilities
as $V_g$ varies (see Fig.\ref{Fig.4}b). The lines are drawn
through the experimental points for visualization. In the range
 of -1.5 V $\le V_g \le$ +3 V, a decrease in the electron density is
accompanied by a marked decrease in their mobility roughly as
$\sim N_s^{3/2}$. Similar dependency of mobility on density is
frequently observed in other two-dimensional structures as well,
where the carrier density is controlled by the electrostatic gate.
It results from the transport time for impurity scattering
depending on the carrier density as $\tau_{tr} \approx
N_s^{\alpha}$ , where $\alpha = 1-2$. In the gate voltage range of
$-2 \le V_g \le -1.5$ V corresponding to the approximate equality
of the electron and hole densities, a sharp jump in the electron
mobility is observed (see the dotted line in Fig.\ref{Fig.4}b). A
further increase in the magnitude of the negative gate bias
slightly reduces the electron mobility and weakly increases the
hole mobility. Of the  most interest is the jump in electron mobility
at $-2 \le V_g \le -1.5$ V. This jump coincides with the gate
voltage range where the hole density first equals and then begins
to exceed the electron density: P$_{s} \ge $ N$_{s}$. We suggest
that this jump may be accounted for by the holes screening and,
therefore, reducing impurity scattering of electrons.

\begin{figure}[h]
\begin{center}\leavevmode
\includegraphics[width=0.9\linewidth]{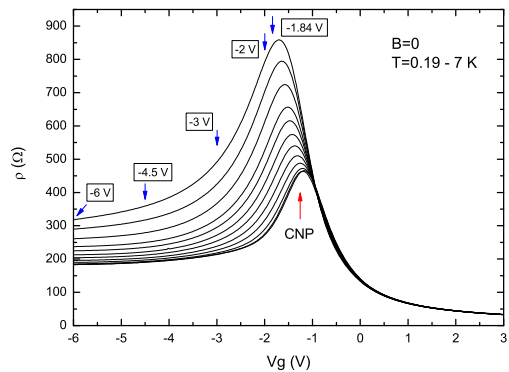}
\caption{Gate voltage dependences $\rho(V_g)$ at $B=0$ and
various temperatures T = (from bottom to top) T= 0.2; 0.5; 1; 1.5;
 2; 2.5; 3; 3.6; 4.1; 5; 6; 7 K. \label{Fig.5}}
\end{center}
\end{figure}

As shown in Section I, in a bipolar system with two types
of charge carriers of the opposite sign, momentum relaxation can be
caused, in addition to other factors, by their mutual scattering
(friction) \cite{levin}. Since only the particles of both kinds
that fall into the $kT$ interval in the vicinity of the Fermi
level are involved in this momentum relaxation mechanism, the
corresponding relaxation time is expected to change with
temperature as $\sim T^{-2}$.

Fig.\ref{Fig.5} presents the gate voltage dependences of our
sample resistance in the zero magnetic field for a number of
temperatures in the interval T = 0.19 - 7 K. Each $\rho(V_g)$
curve has a pronounced maximum. At the lowest temperature T = 0.19
K, the position of this maximum almost coincides with the gate
voltage at which the hole and electron densities are equal (CNP).
Another interesting feature of the curves in Fig.\ref{Fig.5} is
their asymmetric temperature dependence with respect to the gate
voltage. One can see that for $V_g \ge -1$ V, i.e. when the
electrons are the only detectable carriers in the system, there is
only a weak dependence of resistance on temperature. Meanwhile, at
$V_g \le -1$ V, i.e. when the holes begin to populate the valence
band, a considerable increase (by a factor of 1.5 - 3) in
resistance is observed as the temperature increases from 0.2 K to
7 K. It is maximal in the range of $-3 V \le V_g \le -1$ V and
decreasing for higher negative gate biases. Also, with  the
temperature increasing  the maximum of the $\rho(V_g)$ curve shifts
by about 0.5 V to negative gate voltages. In the following section
we analyze the observed behavior using the theory of
electron--hole scattering developed in Section I.

\section{Theory {\it versus} experiment}

Due to the electron--hole scattering, there would be a stronger
temperature dependence of the resistivity in the gate voltage
range where both holes and electrons are present $V_g \le -1$ V
compared to $V_g \ge -1$ V, where the electrons are the only charge
carriers (see Fig.\ref{Fig.4}a). To analyze the system behavior at
$V_g \le -1$ V, we use Eq.(\ref{deltarho}) from Section I,
obtained for the temperature dependence of resistance in a system
with two types of charge carriers when the momentum relaxation is due
to their mutual scattering, which we rewrite in the following
form:

\begin{equation}\label{my_eq}
\rho(T)=\rho_0 \frac{1+(\eta /e)(N_s \mu _p +P_s \mu _n )}{1+(\eta
/e)(N_s -P_s )^2\mu _n \mu _p /(N_s \mu _n +P_s \mu _p )}
\end{equation}

Here, $\rho_0$, $N_s$, $P_s$, $\mu_n$, and $\mu_p$ are the system
resistance, the electron and hole densities and mobilities at $T =
0$, respectively and $\eta$ is the electron-–hole friction
coefficient defined in Sec.I. Irrespective of the details of the
scattering mechanism, at specified values of the electron and hole
densities the probability of electron--hole scattering decreases
as the square of temperature, $\eta=\Theta\cdot T^2$,   where
$\Theta$ is a certain $T$-independent function of $N_s$ and $P_s$
to be determined.

\begin{figure}[h]
\begin{center}\leavevmode
\includegraphics[width=0.9\linewidth]{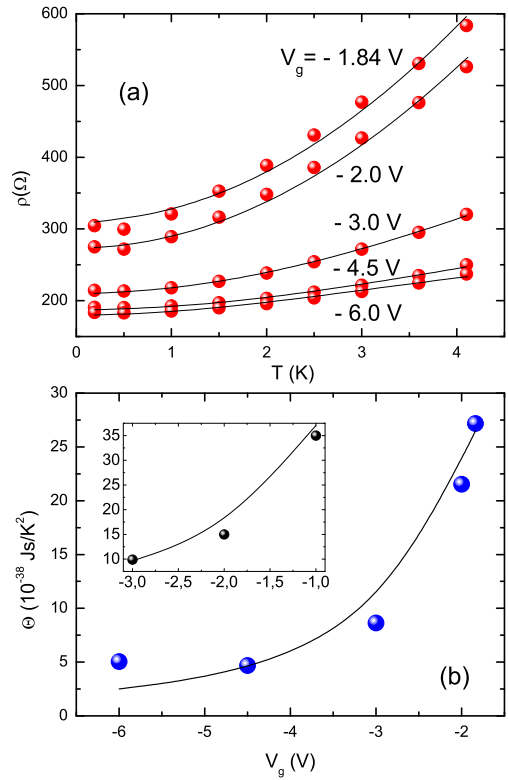}
\caption{(a)-Temperature dependences $\rho(T)$ obtained from
Fig.\ref{Fig.5} for $V_g = –1.84, –2, –3, -4.5, -6 $ V (marked with
arrows in Fig.\ref{Fig.5}). The lines are the fitting by
Eq.(\ref{my_eq}) (see the text); (b)-Parameter $\Theta$ obtained
from fitting the experimental data in Fig.\ref{Fig.5} with
Eq.(\ref{my_eq}). The solid line represents the theory given by
 Eqs.(\ref{etai}-\ref{f1}) with $\alpha=1.2$ and the contact
 interaction constant $\Lambda=1.36$. Insert: similar experimental
 data from another sample published previously \cite{e-h_scatt},
 the line corresponds to Eqs.(\ref{etai}-\ref{f1}) with $\alpha=1.2$
 and the contact interaction constant $\Lambda=1.64$.
\label{Fig.6}}
\end{center}
\end{figure}

In Fig.\ref{Fig.6}a we plot with closed circles the $\rho(T)$
dependences obtained from the experimental curves in
Fig.\ref{Fig.5} at $V_g =-1.84;-2;-3;-4.5;-6$ V (in
Fig.\ref{Fig.5} these gate voltages are marked with arrows). For all
these gate voltages, the resistivity temperature dependence
saturates at $T\le 0.5K$. This allows us to use the values of the
electron and hole mobilities and densities at these temperatures
as zero-T quantities in Eq.(\ref{my_eq}) when fitting it to the
experimental data in Fig.\ref{Fig.6}a. For each of the specified
gate voltages these zero-T parameters were independently obtained
from the magnetotransport data as described in the above
discussion of the curves in Fig.\ref{Fig.2}a,b,c. Therefore, the
fitting procedure for each value of the gate voltage in
Fig.\ref{Fig.6}a depends on a single parameter $\Theta$ in the
expression for $\eta $. The fitting of Eq.(\ref{my_eq}) to the
data is shown in Fig.\ref{Fig.6}a by lines. The temperature range
for fitting was chosen as 0.19-4.1 K. It was found that
Eq.(\ref{my_eq}) does not fit well the experimental points for the
temperatures higher than 4.1 K possibly because of other
temperature dependent scattering mechanisms emerging at these
temperatures. The points in Fig.\ref{Fig.6}b show  fitting
parameter $\Theta$ as a function of the gate voltage. In the
insert we show, for statistics, a similar data for another sample
that was published previously \cite{e-h_scatt}.

\begin{figure}[h]
\begin{center}\leavevmode
\includegraphics[width=0.9\linewidth]{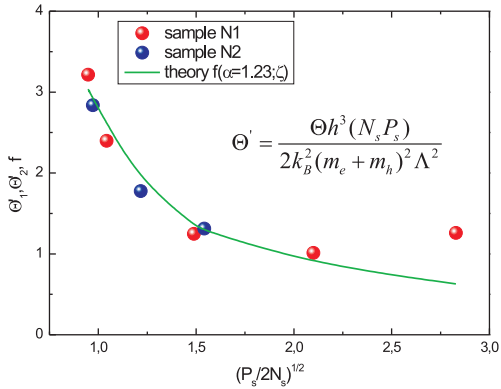}
\caption{Dimensionless quantity $\frac{\Theta
h^3N_sP_s}{2k_{B}^2(m_e+m_h)^2 \Lambda^2}$ plotted as a function
of $\zeta=\sqrt{P_s/2N_s}$ for the two samples data in
Fig.\ref{Fig.6}b. The solid line represents the theory,
Eq.(\ref{f1}). \label{Fig.7}}
\end{center}
\end{figure}

Let us now apply the theoretical results obtained in Sec.I for the
analysis of the gate voltage dependence of $\Theta=\eta/T^2$.
Before we begin, it is vital to note that parameter $\Theta$ gives
a first hand information about the interparticle interaction which
makes our situation rather unique.  Indeed, in the general case of
a 2D electron system with $\sigma \gg e^2/h$ this information can
only be obtained form the study of quantum corrections  which,
apart from being only few percent of the total conductivity depend
on the interaction in an indirect and complicated  form
\cite{altshuler,finkelstein,aleiner}.

First of all we notice that the weak e-h interaction approximation
of our theory, given by Eq.(\ref{tauhe}), does not seem to be
applicable in our case. Indeed, a direct calculation of $\Theta$
using Eq.(\ref{tauhe}) for our system parameters and the
corresponding values of $N_s$, $P_s$ yields $\Theta$ about two
orders of magnitude less than that observed experimentally
(Fig.\ref{Fig.6}b). The reason for this is probably related to the
following fact. In the carrier density range investigated, the
ratio of the screening constant to the Fermi wave vector
$\kappa/\mbox{min}(p_{Fh},p_{Fe})\approx 20 \gg 1$ in which case
treating the e-h interaction as weak becomes unjustified. Besides,
$1/\kappa=1.3$ nm $\ll$ d, where d$=20$ nm is the QW width,  and the
2D consideration loses applicability.

We have taken into theoretical consideration a variety of factors
affecting the e-h interaction, except for its large strength. Besides
the general difficulties associated with the consideration of
strong interaction in a simple 2D case, there are complications due
to the specificity of HgTe quantum wells.

In fact, the individual electron energy levels and wave functions
in a narrow-gap semiconductor are obtained from the size
quantization of a many-component wave function, that results in a
complicated space dependence of electron density. Owing to the
large strength of e-h interaction, its essential part is
accumulated on distances comparable to the well width. This factor
strongly modifies the Coulomb interaction. The long-range
components of the Coulomb interaction are suppressed, while a
short-range 2D scattering amplitude is formed on the scale of the
well width.

 Under these
circumstances we can consider the e-e scattering in the simplest
way, and turn to the short-range potential model represented by
Eq.(\ref{tauhe11}) for the isotropic case and by its extension
(Eqs.(\ref{etai}-\ref{f1})) for the anisotropic spectrum. Then we have
only a single fitting parameter  $\Lambda$, the value of
which can not be found in the 2D model developed here.

The solid lines in Fig.\ref{Fig.6}b and the Insert are the fitting
of Eqs.(\ref{etai}-\ref{f1}) to the experimental dependences of
$\Theta(V_g)$ with $\Lambda=1.36$ (1.64 for the data in the
Insert) and the hole mass anisotropy coefficient $\alpha=1.2$ in
both cases. A more universal way to present the data is to plot
 quantity $\frac{\Theta h^3N_sP_s}{2k_{B}^2(m_e+m_h)^2
\Lambda^2}$ as a function of $\zeta=\sqrt{P_s/2N_s}$, in which
case the experimental points for both of the samples should fall
on the same curve $f(\alpha,\zeta)$. As one can see in
Fig.\ref{Fig.7}, this is indeed the case.

The  values of $\Lambda$ obtained from the fitting in
Fig.\ref{Fig.6}b appear to be too large if we assume that it
represents Coulomb interaction with the dielectric constant of
HgTe equal to 12-15. At the same time, these values are in good
agreement with the short-range model considerations.

Let us discuss the origin of short-range interaction in more
detail. In quantum wells of conventional semiconductors the
subbands are formed from the simple envelope-function states. On
the contrary, in a HgTe quantum layer the size  quantization and
the formation of a gap occur simultaneously. The e-h interaction
modifies the bands, and the value of the effective e-h interaction
is inevitably related to the structure of the states. This
determines the characteristic energy and spatial scales of the e-h
interaction, namely, the gap as the characteristic energy scale
and the width of the quantum layer as the length scale. That
results in a value of $\Lambda$ comaparble with the extracted from
the experimental data.

It should be emphasized, that a large e-e interaction constant
means the inapplicability of the Born approximation for the
electron--hole pair scattering and of the Fermi gas concept.
However,  constant $\Lambda$ can be treated as a low-energy
limit of a dimensionless scattering amplitude that preserves the
above-mentioned estimates. In fact our results constitute an
evidence that even in the case of $\sigma \gg e^2/h$ a 2D e-h
system in HgTe QW should be considered as a strongly correlated 2D
e-h liquid rather than a 2D e-h gas.

\section{Conclusions}
We have developed the theory of temperature-dependent corrections
to the conductivity and magnetotransport coefficients in a 2D
semimetal. These corrections are caused by friction between
electrons and holes.  The corrections obey the quadratic
temperature dependence at the low temperature limit. The friction
coefficients are found for the linear-screened Coulomb
electron--hole interaction and the real spatial structure of the
system. Besides, calculations have been made for the core
electron--hole scattering in the assumption that the Coulomb
potential is completely screened. The experiments were performed
in  20 nm (013) HgTe QW. We found that the conductivity variation
with temperature due to electron--hole scattering is very large
(2-3 times higher than the conductivity in the zero temperature
limit). This gives us the opportunity to obtain a direct
information about interparticle interaction in a 2D electron--hole
system with a high ($\sigma \gg e^2/h$) conductivity value. It has
proved to be impossible to explain the observed strong temperature
dependent variation of conductivity as a consequence of
electron--hole scattering due to Coulomb interaction. Instead, the
short-range e-h scattering model was found to be satisfactory for
the explanation of the observed large friction strength.

\end{document}